\documentclass[
    ,final            
  ]
  {aipproc}

\layoutstyle{6x9}

\usepackage{graphicx} 
\usepackage{psfrag} 
\usepackage{subfigure} 
\usepackage{array} 
\usepackage{booktabs} 
\usepackage{amsfonts} 
\usepackage{amssymb} 
\usepackage{amsmath} 
\usepackage{latexsym} 
\usepackage{amsthm} 
\usepackage{slashed} 
\usepackage{bm} 
\usepackage{mathrsfs} 
\usepackage[usenames,dvipsnames]{color} 
\usepackage{subfigure}

\newcommand{\be}{\begin{equation}}
\newcommand{\ee}{\end{equation}}

\begin{document}

\title{Two-Higgs-doublet models \\ with Minimal Flavour Violation}


\author{Maria Valentina Carlucci}{
  address={Physik-Department, Technische Universit\"at M\"unchen, \\ James-Franck-Stra{\ss}e, D-85748 Garching, Germany}
}

\begin{abstract}
The tree-level flavour-changing neutral currents in the two-Higgs-doublet models can be suppressed by protecting the breaking of either flavour or flavour-blind symmetries, but only the first choice, implemented by the application of the Minimal Flavour Violation hypothesis, is stable under quantum corrections. Moreover, a two-Higgs-doublet model with Minimal Flavour Violation enriched with flavour-blind phases can explain the anomalies recently found in the $\Delta F = 2$ transitions, namely the large CP-violating phase in $B_s$ mixing and the tension between $\epsilon_K$ and $S_{\psi K_S}$.
\end{abstract}

\maketitle


\section{Introduction}

It is well known that the choice of introducing only one Higgs doublet in the Standard Model is just the most economical, but not the only possible one. There is a number of motivations for considering extended Higgs sectors \cite{Gupta:2009wn,Botella:2009pq}; first of all, some models (for example, the Minimal Supersymmetric Standard Model, or some approaches of string theory), require the presence of two or more Higgs doublets; moreover, multi-Higgs models bring many interesting phenomenological features, such as new sources of CP violation, dark matter candidates, axion phenomenology; finally, one should keep in mind that the physical particle spectrum of the Higgs sector (if it exists) is speculation at present and, at the beginning of the LHC era, we should be prepared for surprises. 

However, in the presence of more than one Higgs doublet, the appearance of tree-level Flavour-Changing Neutral Currents (FCNCs) is not automatically forbidden by the standard assignment of the $SU(2)_L\otimes U(1)_Y$ fermion charges as in the Standard Model: additional conditions have to be imposed in order to guarantee a sufficient suppression of FCNC processes. We compare the effectiveness of the two most used mechanisms in the Two-Higgs-Doublet Models (2HDM): Natural Flavour Consevation (NFC) \cite{Glashow:1976nt} and Minimal Flavour Violation (MFV) \cite{Buras:2000dm,D'Ambrosio:2002ex}. We will see that beyond the tree level some FCNCs are naturally generated in both cases, but we will show \cite{Buras:2010mh} that while NFC is not stable under quantum corrections, the renormalization-group invariance of the MFV structure guarantees an adequate suppression at all energies.

The phenomenological tests of these hypotheses are developed on the basis of the FCNC observables of the meson-antimeson mixing. This is an analysis of particular interest, since three interesting problems have recently emerged:
\begin{itemize}
\item first results from CDF \cite{Aaltonen:2007he} and D0 \cite{Abazov:2008fj} indicate a value for the weak phase in the $B_s - \bar B_s$ mixing that is roughly larger than the SM prediction by a factor 20 (but some recent results are smaller);
\item the value of $\sin 2\beta$ resulting from the UT fits tends to be significantly larger than the measured value of $S_{\psi K_S}$~\cite{Buras:2008nn};
\item the values of $S_{\psi K_S}$ and $\epsilon_K$ cannot be 
simultaneously described within the SM~\cite{Buras:2008nn,Lunghi:2008aa}.
\end{itemize}
By decoupling the flavour breaking from the CP violation, we have found that  in the context of a 2HDM with MFV and flavour-blind phases a large phase in the $B_s$ mixing can be easily accommodated, but also that, once this first problem is addressed, the other two issues are automatically solved \cite{Buras:2010mh}.

\section{The Model}

The Higgs Lagrangian of a generic  model with two-Higgs doublets, $H_1$ and 
$H_2$, with hypercharges $Y=1/2$ and $Y=-1/2$ respectively, can be written as
\be
\mathcal{L} = \sum_{i=1,2} D_\mu H_i D^\mu H_i^\dagger + 
\mathcal{L}_{Y}  - V(H_1,H_2)~,
\ee
where $D_\mu H_i = \partial_{\mu}H_i-i g^\prime Y \hat{B}_{\mu}H_i -ig T_a\hat{W^a}_{\mu} H_i$, with $T_a=\tau_a/2$. 
The potential $V(H_1,H_2)$ is such that the $H_i$ gets vacuum expectation value $\langle H^0_{1(2)} \rangle = v_{1(2)}$ with $v = \sqrt{ v_1^2 + v_2^2 } \approx 246 \text{ GeV}$ fixed by the mass of the $W$ boson; moreover, we only consider the case in which it does not contain new sources of CP violation. The Higgs spectrum contains three Goldstone bosons $G^{\pm}$ and $G^0$, two charged Higges $H^{\pm}$, and three neutral Higgses $h^0$ and $H^0$ (CP-even), and $A^0$ (CP-odd).

The most general renormalizable and gauge-invariant interaction of the two Higgs doublets with the SM quarks is 
\be
- \mathcal{L}_Y = \bar Q_L X_{d1} D_R H_1 + \bar Q_L X_{u1} U_R H_1^c 
+ \bar Q_L X_{d2} D_R H_2^c + \bar Q_L X_{u2} U_R H_2 +{\rm h.c.}~,
\ee
where $H_{1(2)}^c = -i\tau_2 H_{1(2)}^*$  and the $X_i$ are $3\times 3$ matrices with a generic flavour structure. By performing a global rotation of angle $\beta = \text{arctan} (v_2/v_1)$ of the Higgs fields $(H_1, H_2)$ to the so-called Higgs basis $(\Phi_v, \Phi_H)$, the mass terms and the interaction terms are separated:
\be
- \mathcal{L}_Y = \bar Q_L \left( \frac{\sqrt{2}}{v} M_{d} \Phi_v + Z_{d} \Phi_H \right)D_R + \bar Q_L \left(\frac{\sqrt{2}}{v} M_{u} \Phi_v^c + Z_{u} \Phi_H^c \right) U_R  +{\rm h.c.}~;
\ee
the quark mass matrices $M_{u,d}$ and the couplings $Z_{u,d}$ are linear combinations of the $X_i$, weighted by the Higgs vacuum expectation values:
\be
M_{u,d} = \frac{v}{\sqrt{2}} \left( \cos \beta X_{{u,d}\,1} + \sin \beta X_{{u,d}\,2} \right)~, \qquad Z_{u,d} = \cos \beta X_{{u,d}\,2} - \sin \beta X_{{u,d}\,1}~.
\ee
In this way it is clear that $M_{u,d}$ and $Z_{u,d}$ cannot be diagonalized simultaneously for generic $X_i$, and we are left with dangerous FCNC couplings to the neutral Higgses.

\section{Protection mechanisms for FCNCs}

A convenient classification of various 2HDMs and of the possible protection of
FCNCs is obtained by identifying how the $X_i$ break the large quark-flavour symmetry of the gauge sector of the SM and the possible continuous or discrete symmetries associated to the Higgs sector. The largest group of unitary quark field transformations that commutes with the SM gauge Lagrangian can be decomposed as \cite{D'Ambrosio:2002ex,Chivukula:1987py}
\be
\mathcal{G}_q = \left(SU(3) \otimes U(1)\right)^3~,
\ee
i.~e. a $SU(3)$ symmetry and a phase symmetry for each electroweak multiplet:
\be
SU(3)^3= SU(3)_{Q_L}\otimes SU(3)_{U_R} \otimes SU(3)_{D_R}~,  \quad U(1)^3 = U(1)_B \otimes  U(1)_Y \otimes U(1)_{\rm PQ}~;
\ee
notice that the three $U(1)$ symmetries can be rearranged as the baryon number, the hypercharge, and the Peccei-Quinn symmetry \cite{Peccei:1977ur}.

One can obtain the suppression of FCNCs by protecting the breaking of one of these two types of symmetry.

\subsection{Natural Flavour Conservation}

It assumes that only one Higgs field can couple to quarks of a given electric charge. This structure can be obtained by imposing appropriate continuous or discrete flavour-blind symmetries.

\begin{itemize}

\item{ \bf Peccei-Quinn symmetry $U(1)_{PQ}$} \\
In this context it can be defined as the symmetry under which $D_R$ and $H_1$ have opposite charges, while all the other fields are neutral. Its application implies
\be
X_{u1}= X_{d2} =0
\ee
and hence eliminates the FCNCs at tree level. However, this symmetry cannot be exact, since it would cause the presence of a massless pseudoscalar Higgs field, and hence it must be broken beyond the tree level; for example, for the down-type Yukawa coupling we can write 
\be
X_{d2} = \epsilon_d \Delta_d
\ee
where $\Delta_d$ is a generic flavour-breaking matrix with $\mathcal{O} (1)$ entries and $\epsilon_d$ that parametrizes the loop suppression, so that we expect $\epsilon_d = \mathcal{O} (10^{-2})$. The comparison with the experimental data can be performed, for example, by considering the CP violation in the system of the $K^0$ mesons, through the observable $\epsilon_K$; the condition $| \epsilon_K^{\text{NP}}| < 0.2 |\epsilon_K^{\rm exp}|$, in the decoupling limit, requires \cite{Buras:2010mh}
\be
| \epsilon_d | \times \left|\text{Im}[( \Delta_{d})^*_{21} ( \Delta_{d})_{12}] 
\right|^{1/2}  \lesssim  3 \times 10^{-7}  \times  \frac{ \cos \beta \; M_H}{100~{\rm GeV}}~,
\ee
i.~e. a large amount of fine-tuning would be needed to provide an efficient protection from FCNCs.

\item{\bf Discrete symmetries $\mathcal{Z}_2$} \\
They are the two discrete subgroups of $U(1)_{PQ}$ under which $H_1 \rightarrow - H_1$ and $D_R~\rightarrow~{\pm}D_R$; they imply  that two of the $X_i$ must vanish: 
\be
X_{u1}= X_{d2} =0~\quad\text{[NFC, Type II]} \quad {\rm or} \quad X_{u2}= X_{d2} =0~\quad \text{[NFC, Type I]}~,
\ee
and hence again a cancellation of the tree-level FCNCs. In principle these symmetries could be exact, but they do not forbid the presence of higher-dimensional operators of the type
\be
\Delta \mathcal{L}_Y = \frac{c_1}{\Lambda^2} \bar Q_L X^{(6)}_{d1} D_R H_1 |H_1|^2 + \frac{c_2}{\Lambda^2} \bar Q_L X^{(6)}_{d2} D_R H_1 |H_2|^2 + \ldots
\ee
that, for the natural values $c_i = \mathcal{O} (1)$ and $\Lambda = \mathcal{O} (1 \text{TeV})$, generate too large FCNCs analogously to the previous case \cite{Buras:2010mh}.

\end{itemize}

\subsection{Minimal Flavour Violation}

It consists of the assumption that the $SU(3)$ quark flavour symmetry is broken only by two independent terms, $Y_d$ and $Y_u$, transforming as
\be
Y_u \sim (3, \bar 3,1)_{SU(3)^3}~, \qquad Y_d \sim (3, 1, \bar 3)_{SU(3)^3}~.
\ee
It implies for the $X_i$ the structure \cite{D'Ambrosio:2002ex}
\begin{subequations}
\begin{align}
X_{d1} &= Y_d \\
X_{d2} &= P_{d2}(Y_u Y_u^\dagger, Y_d Y_d^\dagger) \times Y_d = \epsilon_{0} Y_d + \epsilon_{1} Y_d  Y_d^\dagger Y_d                    
+  \epsilon_{2}  Y_u Y_u^\dagger Y_d + \ldots \\
X_{u1} &= P_{u1}(Y_u Y_u^\dagger, Y_d Y_d^\dagger) \times Y_u = \epsilon^\prime_{0} Y_u + \epsilon^\prime_{1}  Y_u Y_u^\dagger Y_u 
+  \epsilon^\prime_{2}  Y_d Y_d^\dagger Y_u + \ldots \\
X_{u2} &= Y_u
\end{align}
\end{subequations}
that is renormalization group invariant. We notice that at the lowest order in $Y_i Y_i^\dagger$ the $X_i$ are aligned, and hence FCNCs are absent \cite{Pich:2009sp}; they are however generated at higher orders. In order to investigate these FCNCs, one can perform an expansion in powers of suppressed off-diagonal CKM elements, so that the effective down-type FCNC interaction can be written as \cite{D'Ambrosio:2002ex}
\be
\mathcal{L}_{\rm MFV}^{\rm FCNC} \propto {\bar d}^i_L \left[
\left( a_0 V^\dagger \lambda_u^2 V + a_1 V^\dagger \lambda_u^2 V  \Delta + a_2  \Delta V^\dagger \lambda_u^2 V \right) \lambda_d \right]_{ij}
 d^j_R~\frac{S_2 + i S_3}{\sqrt{2}} \; + \text{ h.c. }~,
\ee
where $\lambda_{u,d} \propto 1/v \; \text{diag} \left( m_{u,d},m_{c,s},m_{t,b} \right)$, $\Delta = \text{diag} \left( 0,0,1 \right)$, and the $a_i$ are parameters naturally of $\mathcal{O} (1)$; this structure already shows a large suppression due to the presence of two off-diagonal CKM elements and the down-type Yukawas. 

Also in this case we can derive constraints on the free parameters by imposing that the new physics contributions must be compatible within errors with the experimental data. We have found the bounds \cite{Buras:2010mh}:
\begin{subequations}
\begin{align}
& |a_0| \tan\beta \frac{v}{M_H } < 18 && \text{from} \; \epsilon_K~, \\
& \sqrt{|(a^*_0+a^*_1)(a_0+a_2)|} \tan\beta \frac{v}{M_H } = 10 && \text{from} \; \Delta M_s~, \\
& \sqrt{|a_0+a_1|} \tan\beta \frac{v}{M_H } < 8.5 && \text{from} \; \text{Br}\left( B_s \rightarrow \mu^+ \mu^- \right)~;
\end{align}
\end{subequations}
as can be noted, these conditions are well compatible and perfectly natural.

\section{MFV with flavour-blind phases}

\enlargethispage*{\baselineskip}
The mechanisms of flavour and CP violation do not necessary need to be related: in MFV the Yukawa matrices are the only sources of flavour breaking, but other sources of CP violation could be present, provided that they are flavour-blind \cite{Kagan:2009bn}. Allowing the FCNC parameters $a_i$ to be complex, we investigate the possibility of generic CP-violating flavour-blind phases in the Higgs sector \cite{Buras:2010mh}.

Considering the $\Delta F = 2$ FCNC transitions mediated by the neutral Higgs bosons, the leading MFV effective Hamiltonians are:
\begin{subequations}
\begin{align}
\mathcal{H}^{|\Delta S|=2} & \propto
 - \frac{|a_0|^2}{  M_H^2} ~\frac{m_s}{v} \frac{m_d}{v} \left[ \left( \frac{m_b}{v} \right)^2 V^*_{ts} V_{td} \right]^2~
 ({\bar s}_R d_L) ({\bar s}_L d_R)  {\rm ~+~h.c.}~, \\
\mathcal{H}^{|\Delta B|=2} & \propto
 - \frac{(a^*_0+a^*_1)(a_0+a_2)}{ M_H^2} ~\frac{m_b}{v} \frac{m_q}{v} \left[ \left( \frac{m_b}{v} \right)^2 V^*_{tb} V_{tq} \right]^2~
 ({\bar b}_R q_L) ({\bar b}_L q_R)  {\rm ~+~h.c.}~,
\end{align}
that show two key properties:
\end{subequations}
\begin{itemize}
\item the impact in $K^0$,  $B_{d}$ and $B_{s}$ mixing amplitudes scales with $m_sm_d$, $m_b m_d$ and $m_b m_s$ respectively, opening the possibility of sizable non-standard contributions to the $B_s$ system  without serious constraints from  $K^0$ and  $B_{d}$ mixing;
\item while the possible flavour-blind phases do not contribute to the $\Delta S=2$ effective Hamiltonian, they could have an impact in the $\Delta B=2$ case, offering the possibility to solve the anomaly in the $B_s$ mixing phase.
\end{itemize}
These Hamiltonians have a direct impact on some crucial observables of the neutral mesons systems $N^0-\bar{N}^0$, namely the mass differences $\Delta M_N$ and the asymmetries $S_f$ in the decays $N^0(\bar{N}^0) \rightarrow f$:
\be
\Delta M_N = \frac{1}{m_N} \left| \left\langle N^0 \right| \mathcal{H} \left| \bar{N}^0 \right\rangle \right|~, \qquad S_f = \sin \left( \arg \left\langle N^0 \right| \mathcal{H} \left| \bar{N}^0 \right\rangle \right)~.
\ee

\begin{figure}[h]
\centering
\includegraphics[width=0.9\textwidth,height=3.4cm]{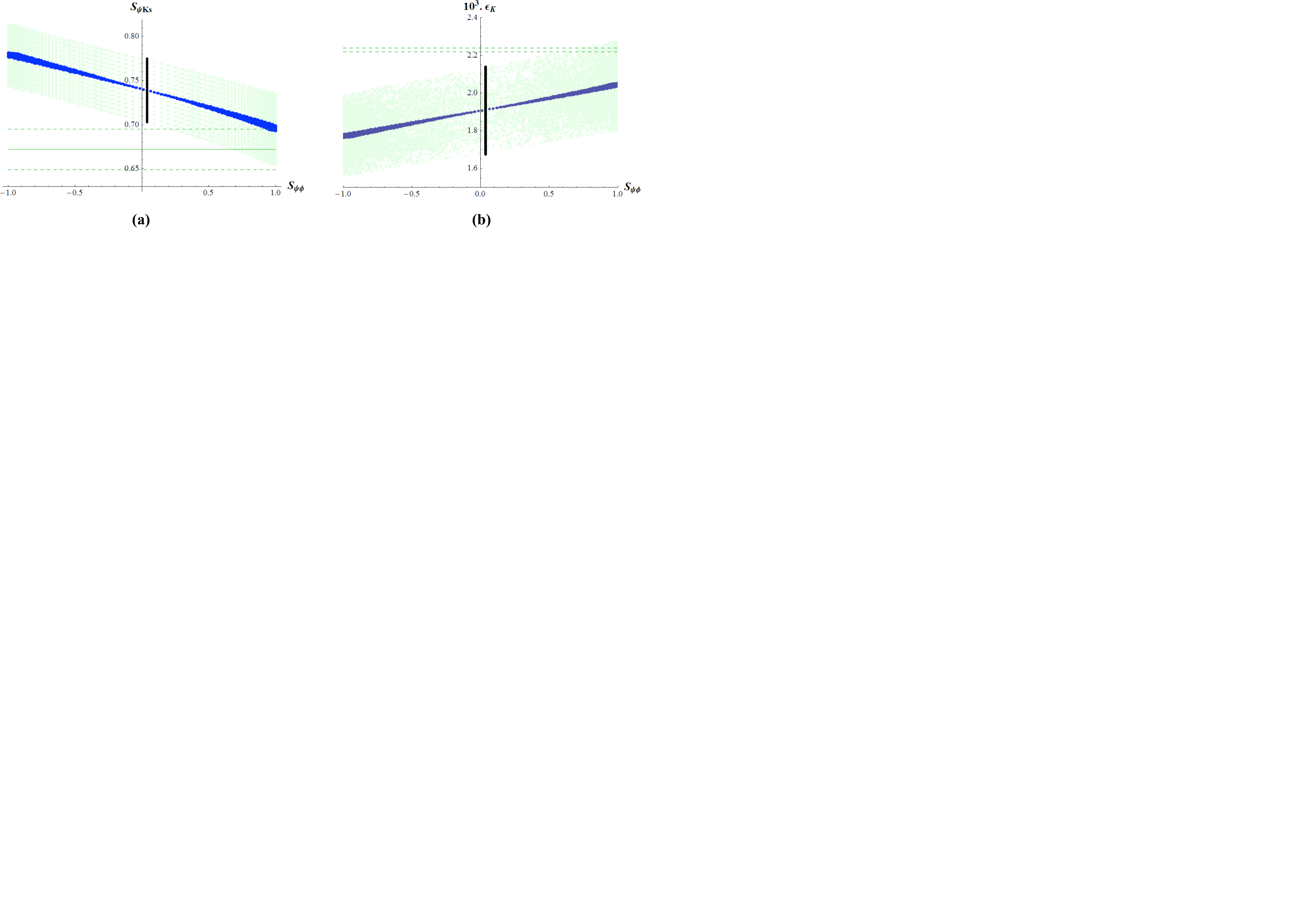}
\caption{{\bf(a)} Correlation between $S_{\psi K_S}$ and $S_{\psi\phi}$. The dark points have been obtained with the CKM phase $\beta$ fixed to its central value: the spread is determined only by the requirement of a deviation of $\Delta M_s$ within 10\% of its SM value; the light points represent the $\pm 1\sigma$ error due to the uncertainty in the extraction of $\beta$. The $\pm1\sigma$ range of $\phi_{B_S}^{\text{exp}}$ (light horizontal lines) and the SM prediction (black vertical line) are also shown. {\bf(b)} Correlation between $\epsilon_K$ and $S_{\psi K_S}$. Notations as before. }
\end{figure}

Using the presence of a new phase in $\mathcal{H}^{|\Delta B|=2}$, we can easily accomodate the hinted large value of $S_{\psi\phi}$ by inserting a new large mixing phase in the $B_s$ system. Moreover, Eq.~16b shows that MFV implies that the new phases in the $B_d$ and the $B_s$ systems are related by the ratio $m_d/m_s$. Hence a large phase in the $B_s$ system determines an unambiguous small shift in the relation between $S_{\psi K_S}$ and the CKM phase $\beta$; as can be seen in Fig.~1a, it goes in the right direction to improve the existing tension between  the experimental value of $S_{\psi K_S}$ and its SM prediction.

Due to the $m_sm_d$ factor in $\mathcal{H}^{|\Delta S|=2}$, the new physics contribution to $\epsilon_K$ is tiny and does not improve alone the agreement between data and prediction for $\epsilon_K$. However, given the modified relation between $S_{\psi K_S}$ and the CKM phase $\beta$, the true value of $\beta$ extracted in this scenario increases with respect to SM fits. As a result of this modified value of $\beta$, also the predicted value for $\epsilon_K$ increases with respect to the SM case, resulting in a better agreement with data (Fig.~1b).


\enlargethispage*{\baselineskip}
\begin{theacknowledgments}
I would like to thank Andrzej J. Buras, Gino Isidori and Stefania Gori for the pleasant and fruitful collaboration; I also would like to thank Marco Bardoscia for technical suggestions and revisions. I am greatly grateful to my first mentor, Prof.~Beppe Nardulli, to whose memory this workshop has been dedicated. This work has been supported in part by the Graduiertenkolleg GRK 1054 of DFG.
\end{theacknowledgments}

%
%
\bibliographystyle{aipproc}   
%

\begin{thebibliography}{99}



\bibitem{Gupta:2009wn}
  R.~S.~Gupta and J.~D.~Wells,
  Phys.\ Rev.\  D {\bf 81} (2010) 055012.
  
\bibitem{Botella:2009pq}
  F.~J.~Botella, G.~C.~Branco and M.~N.~Rebelo,
  Phys.\ Lett.\  B {\bf 687} (2010) 194.
  
\bibitem{Glashow:1976nt}
  S.~L.~Glashow and S.~Weinberg,
  Phys.\ Rev.\  D {\bf 15} (1977) 1958.

\bibitem{Buras:2000dm}
  A.~J.~Buras, P.~Gambino, M.~Gorbahn, S.~Jager and L.~Silvestrini,
  Phys.\ Lett.\  B {\bf 500} (2001) 161.

\bibitem{D'Ambrosio:2002ex}
  G.~D'Ambrosio, G.~F.~Giudice, G.~Isidori and A.~Strumia,
  Nucl.\ Phys.\  B {\bf 645} (2002) 155.

\bibitem{Buras:2010mh}
  A.~J.~Buras, M.~V.~Carlucci, S.~Gori and G.~Isidori,
  arXiv:1005.5310 [hep-ph].

\bibitem{Aaltonen:2007he}
  T.~Aaltonen {\it et al.}  [CDF Collaboration],
  Phys.\ Rev.\ Lett.\  {\bf 100} (2008) 161802.

\bibitem{Abazov:2008fj}
  V.~M.~Abazov {\it et al.}  [D0 Collaboration],
  Phys.\ Rev.\ Lett.\  {\bf 101} (2008) 241801, and arXiv:1005.2757.

\bibitem{Buras:2008nn}
  A.~J.~Buras and D.~Guadagnoli,
  Phys.\ Rev.\  D {\bf 78}, 033005 (2008).
  
\bibitem{Lunghi:2008aa}
  E.~Lunghi and A.~Soni,
  Phys.\ Lett.\  B {\bf 666} (2008) 162.

\bibitem{Chivukula:1987py}
  R.~S.~Chivukula and H.~Georgi,
  Phys.\ Lett.\  B {\bf 188} (1987) 99.

\bibitem{Peccei:1977ur}
  R.~D.~Peccei and H.~R.~Quinn,
  Phys.\ Rev.\  D {\bf 16} (1977) 1791.
  
\bibitem{Pich:2009sp}
  A.~Pich and P.~Tuzon,
  Phys.\ Rev.\  D {\bf 80} (2009) 091702.

\bibitem{Kagan:2009bn}
  A.~L.~Kagan, G.~Perez, T.~Volansky and J.~Zupan,
  Phys.\ Rev.\  D {\bf 80} (2009) 076002.






\end{thebibliography}
%
%


\end{document}